\documentclass[11pt]{article}
\usepackage{amsmath}
\usepackage{amssymb}
\usepackage[mathscr]{eucal}
\usepackage{graphicx, subfigure}
\usepackage{caption}
\usepackage{graphicx}
\renewcommand{\paragraph}{\roman{paragraph}}
\DeclareMathOperator{\sgn}{sgn}
\def \E{\mathbf{E}}
\def \e{\epsilon}

\newtheorem{theorem}{\scshape \mdseries  Theorem}[section]
\newtheorem{lemma}[theorem]{\scshape \mdseries  Lemma}
\newtheorem{coro}[theorem]{\scshape \mdseries  Corollary}

\begin{document}

\title{\sf Compressed Sensing Based on Random Symmetric Bernoulli Matrix\thanks{
Supported by National Natural Science Foundation of China (11071002), Program for New Century Excellent
Talents in University, Key Project of Chinese Ministry of Education
(210091), Specialized Research Fund for the Doctoral Program of
Higher Education (20103401110002), Science and Technological Fund of
Anhui Province for Outstanding Youth (10040606Y33),
Scientific Research Fund for Fostering Distinguished Young Scholars of Anhui
University(KJJQ1001), Academic Innovation Team of Anhui University Project (KJTD001B).}}
\author{Yi-Zheng Fan\thanks{Corresponding author.
 E-mail addresses: fanyz@ahu.edu.cn(Y.-Z. Fan), huangtaooo@yahoo.com.cn (T. Huang), zhu\_m@163.com (M. Zhu)}, Tao Huang, Ming Zhu\\
  {\small  \it $1.$ Key Laboratory of Intelligent Computing and Signal Processing of Ministry of Education,}\\
{\small \it Anhui University, Hefei 230039, P. R. China} \\
  {\small  \it $2.$ School of Mathematical Sciences, Anhui University, Hefei 230601, P. R. China} \\
 }
\date{}
\maketitle

\noindent
{\bf Abstract}\  \
The task of compressed sensing is to recover a sparse vector from a small number of linear and non-adaptive measurements,
and the problem of finding a suitable measurement matrix is very important in this field.
While most recent works focused on random matrices with entries drawn independently from certain probability distributions,
in this paper we show that a partial random symmetric Bernoulli matrix whose entries are not independent,
can be used to recover signal from observations successfully with high probability.
The experimental results also show that the proposed matrix is a suitable measurement matrix.


\newpage

\section{Introduction}
The problem of sparse recovery can be traced back to earlier papers from 90s such as \cite{donoho1,donoho2,donoho3}.
In 2006 the area of compressed sensing made great progress by two ground breaking papers, namely \cite{candes} by Cand\`es, Romberg and Tao and \cite{donoho4} by Donoho.
The Compressed Sensing problem is: Recover $x$ from knowledge of $y=\Phi x$ where $\Phi$ is a suitable  $n \times N$ measurement matrix and $n < N$.
Compressed sensing introduces the extra assumption that the arbitrary vector $x=(x_i)_{i=1}^n\in \mathbb{R}^n$ is {\it $k$-sparse},
 if  the number of non-zero coefficients of vector $x$, denoted by $\|x\|_0:=\#\{i:x_i\neq0\}$, is at most $k$.
 More generally, we assume that $x$ is well-approximated by a sparse vector.
 This discovery has a number of potential applications in signal processing, as well as other areas of science and technology.

It is well known now the question can be solved by $\ell_0$-minimization:
$$\min\|x\|_0    \mbox{~~subject to~~}   y=\Phi x. \eqno(1.1)$$
Considering the difficulties of  this combinatorial optimization problem, actually we solve instead the convex problem:
$$\min\|x\|_1     \mbox{~~subject to~~}    y=\Phi x, \eqno(1.2)$$
where the $\ell_p$-norm is defined  $\|x\|_p=(\sum_{j=1}^n |x_j|^{p})^{1/p} $ as usual.

The matrix $\Phi$ is said to have the {\it Restricted Isometry Property} (RIP) of order $k$ if there exists a $\delta_k \in (0,1)$ such that
$$(1-\delta_k) \|x\|_2^2\leq \|\Phi x\|_2^2\leq (1+\delta_k) \|x\|_2^2\eqno(1.3)$$
for all $k$-sparse vectors $x$. Here $\delta_k$ is  the isometry constant of the matrix $\Phi$, the smallest number satisfied RIP.
Due to \cite{candes,can,donoho4,dono5,lai, gri,rau2} et al,
the $\ell_0$ and $\ell_1$ problems are in fact formally equivalent.
Actually, if $\delta_{2k}<\sqrt{2}-1$, the $\ell_0$ problem has an unique $k$-sparse solution and the solution to the $\ell_1$ problem is that to the $\ell_0$ problem.
In other words, the convex relaxation is exact.
It has been shown that the solution $x^*$ of (1.2) recovers $x$ exactly provided that:
(1) $x$ is sufficiently sparse and (2) the measurement matrix $\Phi $ holds RIP.

The problem that how to choose a suitable measurement matrix $\Phi$ must be investigated in this field.
Most of them are random matrices such as Gaussian or Bernoulli random matrices as well as partial Fourier matrices; see \cite{candes3,rau3,rude}.
It is known \cite{bar,can} that random Gaussian or Bernoulli matrices,
i.e. $n \times N$ matrices with independent and normal distributed or Bernoulli distributed entries satisfy RIP with
probability at least $1-\varepsilon$ provided $k\leq C_1n\log(N/k)+C_2\log\varepsilon^{-1}$, where $C_1$ and $C_2$ are constants depending only on $\delta_k$.
Although Gaussian random matrices are optimal for sparse recovery, they have limited use in practice because many measurement technologies impose structure on the matrix.

Recently the restricted isometry constants of a random Toeplitz type or circulant matrix was estimated,
where the entries of the vector used to generate the Toeplitz or circulant matrices are chosen at random
according to a suitable probability distribution, which are allowed for providing recovery guarantees for $\ell_1$-minimization; see  \cite{raz,pfa,hol,raz2,wakin}.
Compared to Bernoulli or Gaussian matrices, random Toeplitz and circulant matrices have the advantage that they require a reduced number of random numbers to be generated.
More importantly, recovery algorithms tend to be more efficient
when the matrix admits a fast matrix-vector multiply.
 Furthermore, they arise naturally in certain applications such as identifying a linear time-invariant system.
 They close the theoretical gap by providing recovery guarantees for $\ell_1$-minimization in connection with circulant and Toeplitz type matrices
 where the necessary number of measurements scales linearly with the sparsity.
  However, their bound is very pessimistic compared to related estimates for Bernoulli, Gaussian or partial Fourier matrices. More precisely, the estimated number of measurements grows with the sparsity squared, while one would rather expect a linear scaling.

Now we considerate an $N \times N$ symmetric matrix whose entries  $r_{ij}$s hold Bernoulli distribution,
i.e. $r_{ij}$ takes $1,-1$ with probability $1/2$ and $r_{ij}$s are independent for $i \le j$.
It also can be  deduced from the adjacent matrix of a random graph which contains an edge with probability $1/2$ between any two vertices (not necessarily different!).
Choose an arbitrary subset $\Theta\subset \{1,2,...N\}$ of cardinality $n<N$,
 and let $R$ be the {\it partial random symmetric Bernoulli matrix} of size $n\times N$, the submatrix obtained from the above matrix by choosing $n$ rows indexed by $\Theta$.
 Without of loss generation, we choose the first $n$ rows.
 Compared with the matrices mentioned above, it has properties of symmetry and few dependent entries in each column,
 namely it  requires  less random numbers to be generated and there are fast matrix multiplication routines that can be exploited in recovery algorithms.

\section{Our contribution}
The main idea of this paper is motivated by \cite {ach}, as well as some techniques. The key point different to \cite {ach} is that,
we show  Lemma 2.1 below is also valid even the entries in partial random symmetric matrix are not independent.
 Hence this matrix satisfies RIP and can be used as a measurement matrix.

Let $A$ be an $N \times m$ matrix each column corresponding an $N$-dimensional vector.
Let $R$ be an $n \times N$  partial random symmetric matrix.
Considering the projection $f$:
$$ f: A \to n^{-1/2} R A=:E.$$
That is, the $i$th column of $A$ is mapped to the $i$th column of $E$;
and $m$ $N$-dimensional vectors are projected as $m$ $n$-dimensional vectors.
Furthermore, we want to the projection preserves the distance almost invariant, i.e.
$$(1-\e)\|u-v\|_2^2 \le \|f(u)-f(v)\|_2^2 \le (1+\e)\|u-v\|_2^2.\eqno(2.1)$$

Let $\alpha$ be a column vector of $A$.
Then $f(\alpha)=\frac{1}{\sqrt{n}} R\alpha$.
As $f$ is linear, we may normalize $\alpha$ such that $\alpha$ is unit.
For convenience in calculation, take $R=\left(r_{ij}/\sqrt{N}\right)$.
Let $R=(r_1^T,r_2^T,\ldots,r_n^T)^T$ be the row decomposition of $R$.
Then $f(\alpha)=\sqrt{\frac{N}{n}}(r_1 \cdot \alpha, \cdots, r_n \cdot \alpha)=:\sqrt{\frac{N}{n}}(Q_1,\cdots,Q_n)$.
One can get:
$$ \E(Q_j)=0, \ \E(Q_j^2)=\frac{1}{N},\  \E(\|f(\alpha)\|^2)=\frac{N}{n} \sum_{j=1}^n \E(Q_j^2)=1.$$
Let $S=\sum_{j=1}^n Q_j^2$. Then $\|f(\alpha)\|_2^2=S \times \frac{N}{n}$, and
$$\Pr[(1-\e)\|\alpha\|_2^2 \le \|f(\alpha)\|_2^2 \le (1+\e)\|\alpha\|_2^2]=\Pr\left[(1-\e)\frac{n}{N} \le S \le (1+\e)\frac{n}{N}\right].$$

\begin{lemma}
$$\E\left(\Pi_{j=1}^n \exp(hQ_j^2)\right) = \left(\E(\exp(hQ_1^2))\right)^n.$$
\end{lemma}

This lemma guarantees the partial random matrix $R$ has a similar property of Bernoulli matrix discussed in \cite {ach}, and it leads to the below conclusions obviously.

\begin{theorem}
$$\Pr\left[S > (1+\e)\frac{n}{N}\right]< \exp\left(-\frac{n}{2}\left(\frac{\e^2}{2}-\frac{\e^3}{3}\right)\right), \
\Pr\left[S < (1-\e)\frac{n}{N}\right]< \exp\left(-\frac{n}{2}\left(\frac{\e^2}{2}-\frac{\e^3}{3}\right)\right).$$
\end{theorem}

\begin{coro}
Given any $\e,\beta>0$, if $n \ge \frac{4+2\beta}{\e^2/2-\e^3/3}\log m$, then
with probability $1-m^{-\beta}$, (2.1) holds for any two columns $u,v$ of $A$.
\end{coro}

\begin{theorem}
For any give $0<\delta<1$, if taking $\Phi(\omega)=n^{-1/2} R$, and taking $n \ge c_1^{-1}k \log(N/k)$, then
RIP (1.3) holds for $\Phi(\omega)$ with the prescribed $\delta$ and order $k$ with probability $\ge 1-2e^{-c_2n}$,
where $c_1,c_2$ depend only on $\delta$.
\end{theorem}

\begin{lemma}{\em \cite{can}}
Assume that $\delta_{2k}<\sqrt{2}-1$. Then the solution $x^*$ to (1.2) obeys
$$\|x^*-x\|_1\le C_0\|x-x_{(k)}\|_1, \; \|x^*-x\|_2 \le C_0 k^{-1/2} \|x-x_{(k)}\|_1,$$
for some constant $C_0$, where $x_{(k)}$ is obtained from $x$ by setting all but the $k$-largest entries to be zero.
In particular if $x$ is $k$-sparse, the recovery is exact.
\end{lemma}

If the measurements are corrupted with noise, that is
$$y = \Phi x +z, \eqno(2.2)$$
where $z$ is an unknown noise term.
We will consider the following problem:
$$\min_{x \in \mathbb{R}^N}\|x\|_1 \mbox{~~subject to~~} \|y-\Phi x\|_2\le \e,\eqno(2.3)$$
where $\e$ is an upper bound on the size of the noisy contribution.

\begin{lemma}{\em \cite{can}}
Assume that $\delta_{2k}<\sqrt{2}-1$ and $\|z\|_2 \le \e$. Then the solution $x^*$ to (2.3) obeys
$$\|x^*-x\|_2 \le C_0 k^{-1/2} \|x-x_{(k)}\|_1+C_1\e,$$
for some constants $C_0,C_1$.
\end{lemma}

So, if we recover a $k$-sparse vector $x$, in Theorem 2.4 taking $\delta$ such that $0<\delta<\sqrt{2}-1$, and $n \ge c_1^{-1}2k \log(N/(2k))$, using the matrix
$n^{-1/2} R$ as $\Phi$, then $\Phi$ obeys RIP with order $2k$ and $\delta<\sqrt{2}-1$.
By Lemma 2.5, with high probability, we could recover $x$ exactly.

\section{Proofs}
{\bf Proof of Lemma 2.1.}
We first prove if taking $B=\{r_{12}=a_2,r_{13}=a_3,\ldots,r_{1n}=a_n\}$ in $\sum_{j=2}^n hQ_j^2$, the expectation
$\E(\exp(\sum_{j=2}^n hQ_j^2)|B)$ is independent of $B$.
\begin{align*}
\E(\exp(\sum_{j=2}^n hQ_j^2)|B)&=\E(\exp(h((\alpha_1 a_2+\sum_{k=2}^N \alpha_k r_{2k})^2+\cdots+(\alpha_1 a_n+\sum_{k=2}^N \alpha_n r_{nk})^2)|B)\\
&=\E(\exp(h((\alpha_1 |a_2|+\sum_{k=2}^N \alpha_k \sgn(a_2)r_{2k})^2+\cdots+(\alpha_1 |a_n|+\sum_{k=2}^N \alpha_n \sgn(a_n) r_{nk})^2)|B)\\
&=\E(\exp(h((\alpha_1 +\sum_{k=2}^N \alpha_k r_{2k})^2+\cdots+(\alpha_1 +\sum_{k=2}^N \alpha_n  r_{nk})^2).
\end{align*}
Observe
\begin{align*}
\E(\exp(\sum_{j=2}^n hQ_j^2))&=\sum_{a_2,a_3,\ldots,a_n}\E(\exp(\sum_{j=2}^n hQ_j^2)|B))\Pr(B)\\
&=\E(\exp(\sum_{j=2}^n hQ_j^2)|B) \sum_{a_2,a_3,\ldots,a_n}\Pr(B)\\
&=\E(\exp(\sum_{j=2}^n hQ_j^2)|B)
\end{align*}
Now we have
\begin{align*}
\E(\Pi_{j=1}^n \exp(hQ_j^2)) &= \sum_{a_2,a_3,\ldots,a_n}\E(\exp(hQ_1^2 \cdot \exp(\sum_{j=2}^n hQ_j^2))|B)\Pr(B)\\
&=\sum_{a_2,a_3,\ldots,a_n}\E(\exp(hQ_1^2)|B)\E(\exp(\sum_{j=2}^n hQ_j^2)|B)\Pr(B)\\
&=\E(\exp(\sum_{j=2}^n hQ_j^2)|B) \sum_{a_2,a_3,\ldots,a_n}\E(\exp(hQ_1^2)|B)\Pr(B)\\
&=\E(\exp(\sum_{j=2}^n hQ_j^2)|B) \E(\exp(hQ_1^2)\\
&=\E(\exp(hQ_1^2)) \E(\exp(\sum_{j=2}^n hQ_j^2)).
\end{align*}
The result holds by induction. \hfill $\blacksquare$

\begin{lemma} {\em \cite{ach}}
For all $h \in [0,N/2)$ and all $N \ge 1$,
$$ \E(\exp(hQ_1^2)) \le \frac{1}{\sqrt{1-2h/N}}\eqno(3.1)$$
$$ \E(Q_1^4)\le \frac{3}{N^2}.\eqno(3.2)$$
\end{lemma}

\noindent{\bf Proof of Theorem 2.2.}
The proof is very similar to that in \cite[Lemma 5]{ach}, combining with Lemmas 2.1 and 3.1.
For arbitrary $h > 0$,  $$\Pr[S>(1+\e)\frac{n}{N}]=\Pr[\exp(hS)>\exp(h(1+\e)\frac{n}{N})]<\E(\exp(hS)\exp(-h(1+\e)\frac{n}{N})).$$
By Lemma 2.1, we get $$\E(\exp(hS)=(\E(\exp(hQ_1^2)))^n.$$
Thus for any $\e>0$,$$\Pr[S>(1+\e)\frac{n}{N}]<(\E(\exp(hQ_1^2)))^n\exp(-h(1+\e)\frac{n}{N})).\eqno(3.3)$$
Similarly, but this time considering $\exp(-hS)$ for arbitrary $h>0$, we get that for any $\e>0$, $$\Pr[S<(1-\e)\frac{n}{N}]<(\E(\exp(-hQ_1^2)))^n\exp(h(1-\e)\frac{n}{N})).\eqno(3.4)$$

Substituting (3.1) in (3.3) we get (3.5). To optimize the bound we set the derivative in (3.5) with respect to $h$ to $0$.
This gives $h=\frac{N}{2}\frac{\e}{1+\e}<\frac{N}{2}$.
Substituting this value of $h$ and series expansion yields (3.6).
$$
\Pr[S>(1+\e)\frac{n}{N}]  \leq(\frac{1}{\sqrt(1-h/N)})^n\exp(-h(1+\e)\frac{n}{N})\eqno(3.5)$$
$$~~~~~~~~~~~~~~~~~~~~~~~~=((1+\e)\exp(-\e))^{n/2}<\exp(-\frac{n}{2}(\e^2/2-\e^3/3))\eqno(3.6)$$
Similarly, substituting (3.2) in (3.4)  and taking $h=\frac{N}{2}\frac{\e}{1+\e}$, we get  $$\Pr[S<(1-\e)\frac{n}{N}]<\exp(-\frac{n}{2}(\e^2/2-\e^3/3))\eqno(3.7)$$
\hfill $\blacksquare$

\noindent{\bf Proof of Corollary 2.3.}
For any column $\alpha$ of $A$, by Theorem 2.2,
 $$\Pr[\|f(\alpha)\|_2^2 < (1-\e)\|\alpha\|_2^2 \mbox{~or~} \|f(\alpha)\|_2^2  > (1+\e)\|\alpha\|_2^2]<2\exp(-\frac{n}{2}(\e^2/2-\e^3/3)).$$
There are ${m} \choose {2}$ pairs of $u,v$ of the columns of $A$.
So, taking $\alpha=u-v$ in the above inequality, we have
 $$\Pr[\|f(u-v)\|_2^2 < (1-\e)\|u-v\|_2^2 \mbox{~or~} \|f(u-v)\|_2^2  > (1+\e)\|u-v\|_2^2, \mbox{~for all~}u,v]<2{{m} \choose {2}}\exp(-\frac{n}{2}(\e^2/2-\e^3/3)).$$
Hence, if $n \ge \frac{4+2\beta}{\e^2/2-\e^3/3}\log m$,
then $$\Pr[(1-\e)\|u-v\|_2^2 \le \|f(u-v)\|_2^2 \le (1+\e)\|u-v\|_2^2 \mbox{~for all~}u,v]>1-m^{-\beta}.$$ \hfill $\blacksquare$

Let $(\Omega,\rho)$ be a probability measure space and let $r$ be a random variable on $\Omega$.
Given $n$ and $N$, we can generate random matrix $\Phi$ by choosing the entries $r_{ij}\;(i=1,\ldots,n;j=1,\ldots,N)$ as (not necessarily independent) realizations of $r$.
This yields the random matrix $\Phi(\omega)$.

If the probability distribution generating the matrix $\Phi(\omega)$ holds the following concentrated inequality:
$$\Pr[|\|\Phi(\omega)x\|_2^2-\|x\|_2^2|\ge \e \|x\|_2^2] \le 2 e^{-nc_0(\e)}, 0<\e<1, \eqno(3.8)$$
where the probability is taken over all $n \times N$ matrices $\Phi(\omega)$ and $c_0(\e)$ is only depending on $\e$ and $c_0(\e)>0$ for all $\e$,
then RIP holds for $\Phi(\omega)$ with high probability; see the following result.

\begin{lemma}{\em \cite{bar}}
Suppose that $n,N$, and $0<\delta<1$ are given.
If $\Phi(\omega)$ satisfies (3.8), then there exists constant $c_1,c_2>0$ depending only on $\delta$ such that
RIP (1.3) holds for $\Phi(\omega)$ with the prescribed $\delta$ and any $k \le c_1n/\log(N/k)$ with probability $\ge 1-2e^{-c_2n}$.
\end{lemma}

{\bf Remark:}
1. In Lemma 3.2, it is valid if taking $k \le c'_1 n / [\log(N/n)+1]$ for $c'_1>0$ only depending on $c_1$.

2. If we need the RIP (1.3) holds with order $k$, we take $n \ge c_1^{-1}k \log(N/k)$.
So, Theorem 2.4 is asserted.

\section{Experiments}
Let $x$ be a $k$-sparse discrete signal with length $256$ whose nonzero entries are $1$ or $-1$.
The sensing matrix $R$ is partial random symmetric Bernoulli matrix.
The classical convex optimization algorithm $\ell_1$-minimization is used for reconstruction.
The experimental results are compared with those of Bernoulli, random Gaussian, Toeplitz and circulant  matrices,
where the entries of Gaussian matrix are chosen from a normal distribution with mean zero and variance one,
    the Toeplitz matrix is generated by the first two rows of the Gaussian matrix, and the circulant matrix is generated by the first row.

   We first analysis the performances of the matrices under different sparsity.
   Set the  measurement number $n=100$.
    The results of $1000$ experiments are summarized in Fig. 4.1, from which
    we see that as the sparsity increases, all the performances  decrease.
    It is hard to distinguish which one is the best among Bernoulli matrix (B), Gaussian matrix (G), Toeplitz matrix (T), Circulated  matrix (C) and  $R$.

 \begin{figure}[!h]
  \centering
  \renewcommand\thefigure{\arabic{section}.\arabic{figure}}
    \begin{minipage}[]{1\textwidth}
      \centering
     \includegraphics[width=1\textwidth]{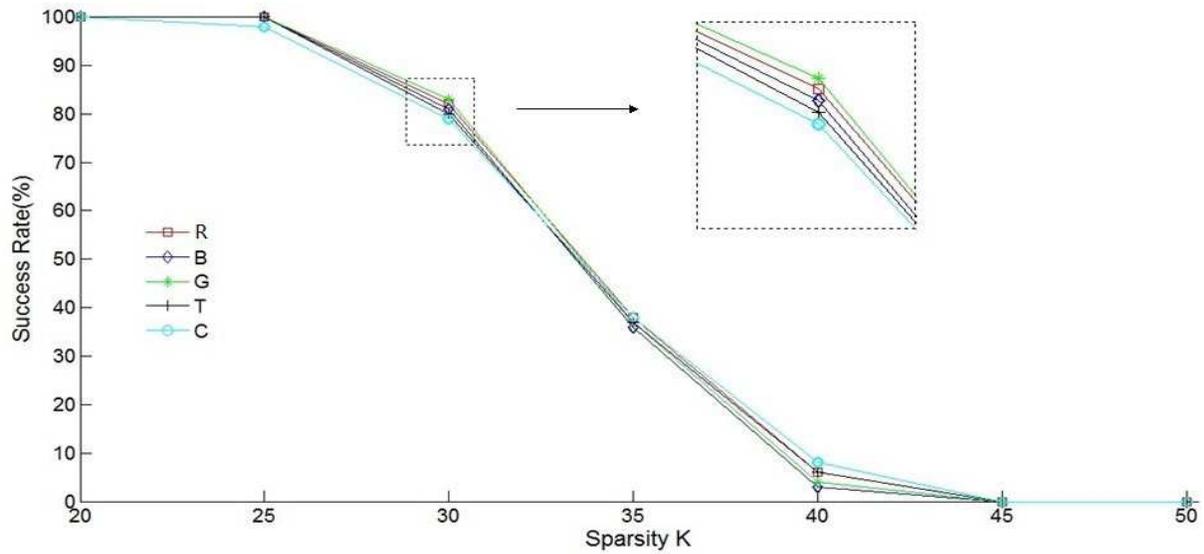}\\
    \caption{ Success rate as a function of sparsity K}
   \end{minipage}
     \hspace{0in}
\end{figure}

   We also investigate the performances of the matrices under different measurement numbers.
   Set the sparsity $k=20$.
   The results of $1000$ experiments are summarized and shown in Fig. 4.2.
   When the measurement number $n$ becomes large, the performance of all matrices get better.
   Especially, when $n \ge 95$ almost all experiments are successful.

 \begin{figure}[!h]
  \centering
  \renewcommand\thefigure{\arabic{section}.\arabic{figure}}
    \begin{minipage}[]{1\textwidth}
      \centering
     \includegraphics[width=1\textwidth]{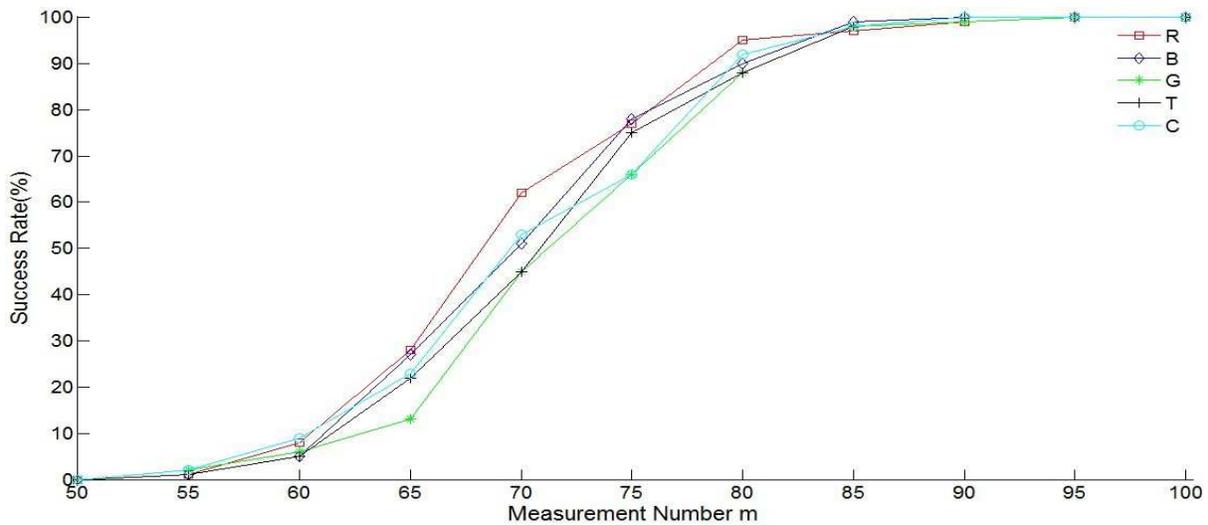}\\
    \caption{Success rate as a function of measurement number m}
   \end{minipage}
     \hspace{0in}
\end{figure}

   Next we check the performances of the above sensing matrices through the real image reconstruction experiment.
   The original image is shown in Fig. 4.3, with size of $64\times 64$ and sparsity $k=739$.
   Set measurement number $n=2400$.
   The mean square error (MSE) is defined as $MSE=\frac{\|X-M\|_F}{\|M\|_F}$,
   where $\|\cdot\|_F$ being the Frobenius norm, $X$ is the reconstruction and $M$ is the original image.
The experimental results are shown in Fig. 4.3.

\begin{figure}[!h]
  \centering
  \renewcommand\thefigure{\arabic{section}.\arabic{figure}}
  \subfigure{
    \begin{minipage}{0.31\textwidth}
      \centering
     \includegraphics[width=0.9\textwidth]{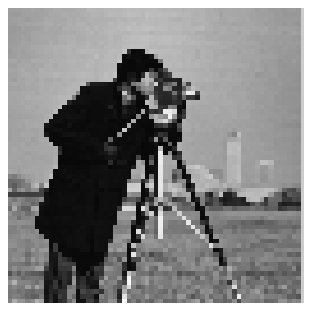}\\
   \caption*{Original image}
  \end{minipage}}
     \hspace{0in}
\subfigure{
   \begin{minipage}{0.31\textwidth}
      \centering
     \includegraphics[width=0.9\textwidth]{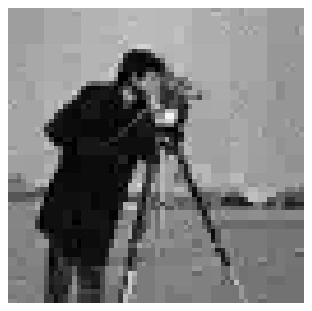}\\
    \caption*{R(MSE=0.0664)}
   \end{minipage}}
     \hspace{0in}
     \subfigure{
   \begin{minipage}{0.31\textwidth}
      \centering
     \includegraphics[width=0.9\textwidth]{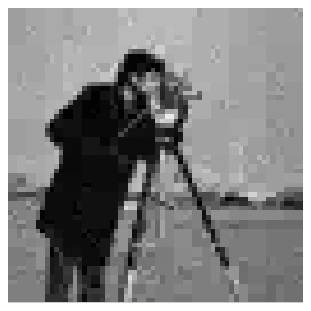}\\
    \caption*{Bernoulli(MSE=0.0672)}
   \end{minipage}}
  \subfigure{
    \begin{minipage}{0.31\textwidth}
      \centering
     \includegraphics[width=0.9\textwidth]{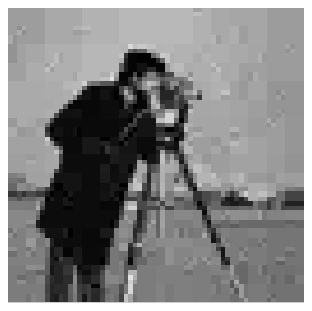}\\
    \caption*{Gaussian (MSE=0.0681)}
   \end{minipage}}
     \hspace{0in}
      \subfigure{
   \begin{minipage}{0.31\textwidth}
      \centering
     \includegraphics[width=0.9\textwidth]{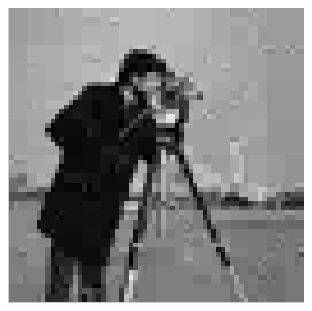}\\
    \caption*{Toeplitz(MSE=0.0647)}
   \end{minipage}}
     \hspace{0in}
      \subfigure{
    \begin{minipage}{0.31\textwidth}
      \centering
     \includegraphics[width=0.9\textwidth]{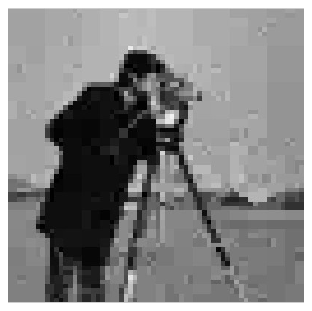}\\
    \caption*{Circulant(MSE=0.0684)}
   \end{minipage}}
     \hspace{0in}
     \caption{ Real world data reconstruction}
\end{figure}

   In practice, the sampled signal usually meets some unavoidable noises.
   As a result, it is necessary to check the performances of our sensing matrix $R$ under different noise levels.
   Gaussian random noise with mean value $0$ and standard deviation whose value is chosen from \{0, 0.2, 0.4, 0.6, 0.8, 1.0\} is added to the measurement value of the image. Experimental results are shown in Fig. 4.4. The increased noise level leads to the poor reconstruction performance.

\begin{figure}[!h]
  \centering
  \renewcommand\thefigure{\arabic{section}.\arabic{figure}}
    \begin{minipage}[]{1\textwidth}
      \centering
     \includegraphics[width=1\textwidth]{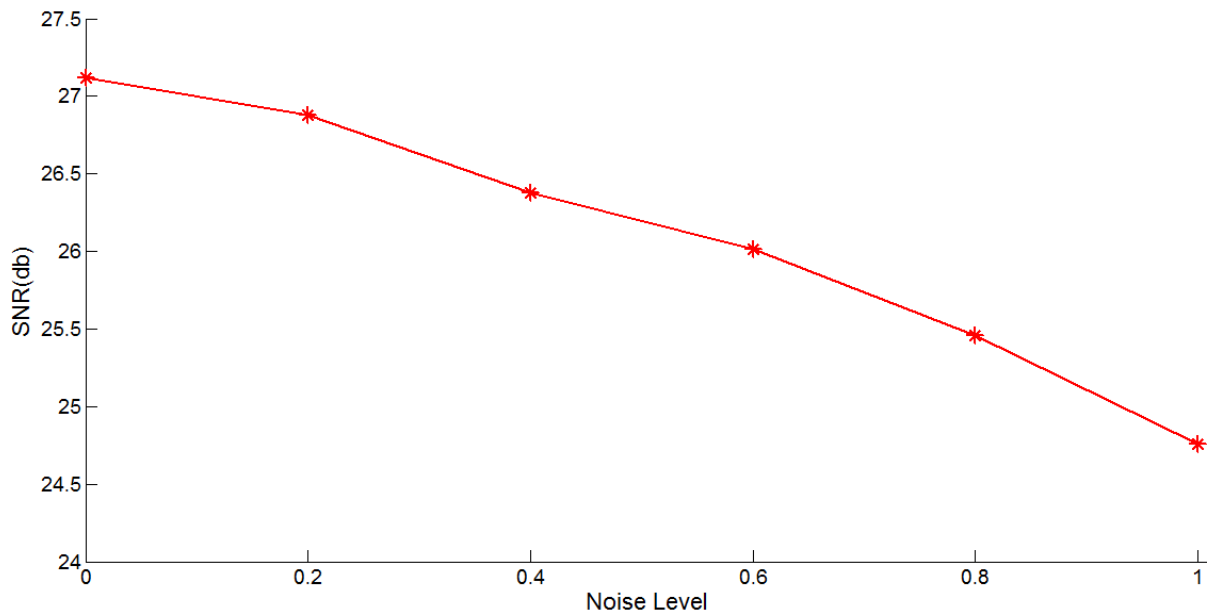}\\
    \caption{Signal Noise Ratio(SNR) under different noise levels}
   \end{minipage}
     \hspace{0in}
\end{figure}

\section{Conclusion}

As we know the equality $\E(XY)=\E X \cdot \E Y$ may hold even if $X,Y$ are not independent.
To a certain extent the partial random symmetric Bernoulli matrix may have the similar properties with Gaussian or Bernoulli matrix.
The theoretical analysis and experiment results show that, we can use this partial random Bernoulli  matrix as the measurement matrix in Compressed Sensing.

Furthermore, there is a relationship between this matrix and random graph.
Recall that the Erd\"os-R\'enyi model $\mathscr{G}_n(p)$ consists of all graphs on $n$ vertices in which the edges are chosen
independently with probability $p\in (0,1)$ (see \cite{bollo}).
If letting $A(G)$ be the adjacency matrix of a graph $G \in \mathscr{G}_n(1/2)$, then
$2A(G)-J$ is a random symmetric matrix whose entries hold Bernoulli distribution, where $J$ is a matrix consisting of all ones.
So it is hopeful to solve some CS problems based on random graphs.
We will seriously considered it in future work.


\small


\begin{thebibliography}{90}

\bibitem{ach} D. Achlioptas, Database-friendly random projections: Johnson-Lindenstrauss with binary coins, {\it J. Comput. System Sci.}, 66(4): 671-687, 2003.

\bibitem{raz} W. Bajwa, J. Haupt, G. Raz, S. Wright, R. Nowak, Toeplitz structured compressed sensing matrices, {\it IEEE/SP Workshop on Statistical Signal Processing-SSP}, 2007.


\bibitem{bar} R. Baraniuk, M. Davenport, R. DeVore, M. Wakin, A simple proof of the restricted isometry property for random matrices,
{\it Constr. Apporx.}, 28(3): 253-263, 2008.

\bibitem{bollo} B. Bollob\'as, {\it Random Graphs (2nd ed.)}, Cambridge University Press, 2001.

\bibitem{candes} E. Cand\`es, J.Romberg, T. Tao, Robust uncertainty principles:Exact signal recostruction from highly incomplete Fourier information,
{\it IEEE Trans. Inform. Theory}, 52(2): 489-509, 2006.

\bibitem{candes3} E. J. Cand\`es,  T. Tao, Near optimal signal recovery from random projections: universal encoding strategies,
{\it IEEE Trans. Inform. Theory}, 52(12): 5406-5425, 2006.


\bibitem{can} E. J. Cand\`es, The restricted isometry property and its implication for compressed sensing, {\it Comptes Rendus Mathematique}, 346(9): 589-592, 2008.

\bibitem{donoho1} D. L. Donoho, P. B. Starck, Uncertainty principles and signal recovery, {\it SIAM J. Appl. Math.}, 49(3): 906-931, 1989.

\bibitem{donoho3} D. L. Donoho, X. Huo, Uncertainty principles and ideal atomic decomposition, {\it IEEE Trans. Inform. Theory}, 47(7): 2845-2862, 2001.

\bibitem{donoho2} D. L. Donoho, M. Elad, Optimally sparse representation in general (nonorthogonal) dictionaries via minimization,
{\it  Proc. Natl. Acad. Sci.-PNAS}, 100(5): 2197-2202, 2003.

\bibitem{donoho4} D. L. Donoho, Compressed sensing, {\it IEEE Trans. Inform. Theory}, 52(4): 1289-1306, 2006.

\bibitem{dono5} D. L. Donoho, J. Tanner, Counting faces of randomly-projected polytopes when the projection radically lowers dimension, {\it J. Amer.
Math. Soc}, 22(1): 1-53, 2009.

\bibitem{git} Y. C. Eldar, G. Kutyniok, {\it Compressed Sensing: Theory and Applications}, Cambridge University Press, 2012.

\bibitem{lai} S. Foucart, M. Lai, Sparsest solutions of underdetermined linear systems via $l_q$-minimization for $0<q\leq 1$, {\it Appl. Comput. Harmon. Anal.},
26(3): 395-407, 2009.

\bibitem{gri} R. Gribonval, M. Nielsen, Sparse representations in unions of bases, {\it IEEE Trans. Inform. Theory}, 49(12): 3320-3325, 2003.

\bibitem{pfa} G. Pfander, H. Rauhut, J. Tropp, The restricted isometry for time-frequency structured random matrices, {\it Probab. Theory Relat. Fields}, doi:  10.1007/s00440-012-0441-4.

\bibitem{rau2} H. Rauhut, Random sampling of sparse trigonometric polynomials, {\it Appl. Comput. Harmon. Anal.}, 22(1): 16-42, 2007.

\bibitem{rau3} H. Rauhut, Stability results for random sampling of sparse trigonometric polynomials, {\it IEEE Trans. Inform. Theory}, 54(12): 5661-5670, 2008.

\bibitem{hol} H. Rauhut, E. Allee, Circulant and Toeplitz Matrices in Compressed Sensing, {\it Computing Research Repository}, vol. abs/0902.4, 2009

\bibitem{rau} H. Rauhut, J. Romberg, J. Tropp, Restricted isometries for partial random circulant matrices, {\it Appl. Comput. Harmonic Anal.}, 32(2): 242-254, 2012.

\bibitem{raz2} J. Romberg, G. Raz, S. Wright, R. Nowak, Compressive sensing by random convolution, {\it  SIAM J. Imaging Sci.}, 2(4): 1098-1128, 2009.

\bibitem{rude} M. Rudelson, R. Vershynin, Sparse reconstruction by convex relaxation: Fourier and Gaussian measurements,
{\it Conference on Information Sciences and Systems-CISS}, 2006.

\bibitem{wakin} J. Tropp, M. Wakin, M. Duarte, D. Baron, R. Baraniuk, Random filters for compressive sampling and reconstruction,
{\it Int. Conf. Acoustics, Speech, and Signal Processing}, vol. 3, pp. III-872-875, 2006.



\end{thebibliography}
\end{document}